# Observation of acoustic Landau quantization and quantum-Hall-like edge states

Xinhua Wen[1], Chunyin Qiu[1*], Yajuan Qi[1], Liping Ye[1], Manzhu Ke[1],

Fan Zhang[2], and Zhengyou Liu[1,3]

[1]Key Laboratory of Artificial Micro- and Nano-structures of Ministry of Education and School of Physics and Technology, Wuhan University, Wuhan 430072, China

[2]Department of Physics, University of Texas at Dallas, Richardson, Texas 75080, USA

[3]Institute for Advanced Studies, Wuhan University, Wuhan 430072, China

**Many intriguing phenomena occur for electrons under strong magnetic fields[1,2]. Recently, it was proposed that an appropriate strain texture in graphene can induce a synthetic gauge field[3-6], in which the electrons behave like in a real magnetic field[7-11]. This opened the door to control quantum transport by mechanical means and to explore unprecedented physics in high-field regime. Such studies have been achieved in molecular[12] and photonic[13] lattices. Here we report the first experimental realization of giant uniform pseudomagnetic field in acoustics by introducing a simple uniaxial deformation to acoustic graphene. Benefited from the controllability of our macroscopic platform, we observe the acoustic Landau levels in frequency-resolved spectroscopy and their spatial localization in pressure-field distributions. We further visualize the quantum-Hall-like edge states (connected to the zeroth Landau level), which have been elusive before owing to the challenge in creating large-area uniform pseudomagnetic fields[5,6]. These results, highly consistent with our full-wave simulations, establish a complete framework for artificial structures under constant pseudomagnetic fields. Our findings, conceptually novel in acoustics, may offer new opportunities to manipulate sound.**

Unlike electrons, neutral particles such as photons and phonons do not carry charges and are inert to an external magnetic field. As a consequence, many peculiar phenomena related to charged particles in magnetic fields are not accessible for neutral particles. However, based on the concept of pseudomagnetic field (PMF), observation of magnetic-field-like effects is feasible in artificial structures that mediate light or sound[13-17]. This offers an unconventional way to control such classical waves. Together with the robustness to temperature and the absence of inter-particle interaction, the exceptional macroscopic controllability allows the artificial platforms to be tractable classical counterparts for probing the tantalizing quantum physics that demands intricate atomic-scale manipulations. Outstanding examples can be referred to the recent observations of topological band phenomena in artificial structures[18-25].

Essentially, the PMF in a strained graphene comes from the spatially dependent momentum shift of Dirac points, the linearly crossing points in energy spectrum at the inequivalent hexagonal Brillouin zone corners $K$ and $K'$. Unless stated otherwise, below we focus on the physics in $K$ valley since all that in $K'$ valley can be obtained by applying time reversal operation. For a strained graphene, the synthetic vector gauge field $\mathbf{A}$ enters the effective Dirac Hamiltonian $\hat{H}_\mathbf{k} = \hbar v_D \boldsymbol{\sigma} \cdot \mathbf{k}$ via the transformation $\hbar \mathbf{k} \to \hbar \mathbf{k} + e\mathbf{A}$ with $\mathbf{A} = \hbar e^{-1} \delta \mathbf{k}$. Here $\hbar$, $e$, and $v_D$ are reduced Planck constant, electron charge, and Dirac velocity, respectively; $\boldsymbol{\sigma} = (\sigma_x, \sigma_y)$ are Pauli matrices, $\mathbf{k} = (k_x, k_y)$ is the momentum measured from the Dirac point, and $\delta \mathbf{k}$ depicts the strain-induced shift of Dirac point. An appropriate spatial distribution of the vector potential, $\mathbf{A}(\mathbf{r})$, achieved usually by imposing a triaxial stretch to the graphene flakes[3,4], can give rise to a uniform PMF $\mathbf{B} = \nabla \times \mathbf{A}$. Similar ideas are also applicable to photonic and mechanical structures with conic dispersions[13-17]. Here we propose a novel strategy to create a uniform PMF for airborne sound: we construct a two-dimensional (2D) sonic crystal (SC) with linearly varying vector potential by simply arranging elliptical scatterers with a gradient shape factor along one direction (Figs. 1a and 1b).

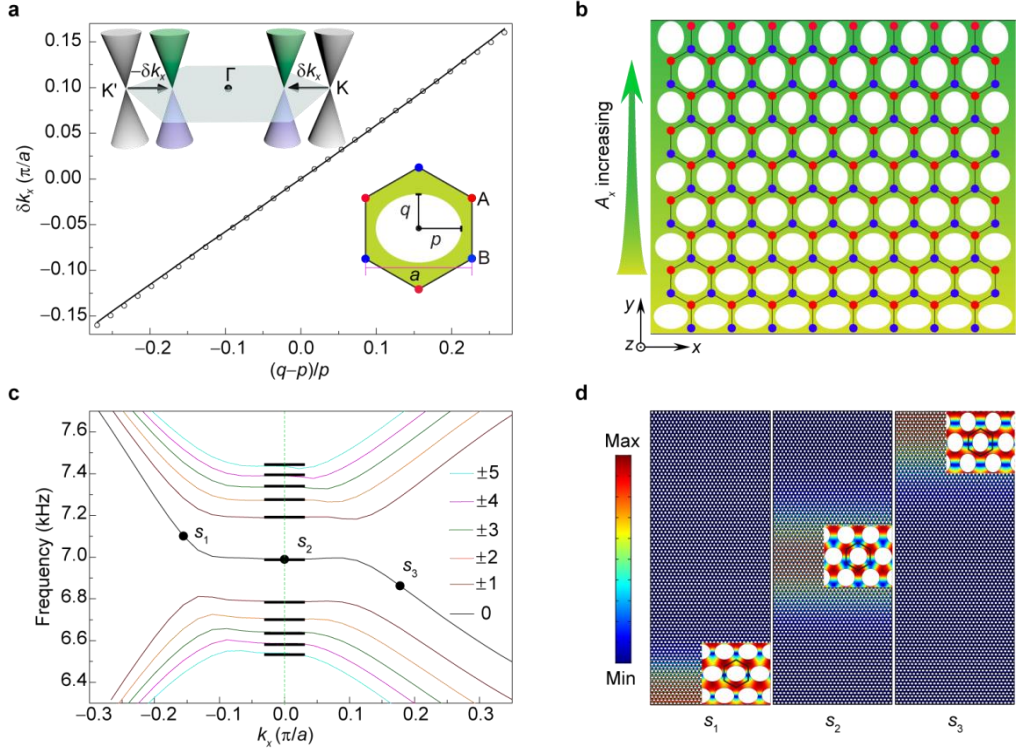

**Figure 1 | Synthesized acoustic magnetic field and relativistic Landau quantization. a**, Simulated Dirac point shifts (circles) in momentum space ($\delta k_x$, illustrated in the upper inset) for the SCs made of uniform elliptical scatterers identified with radii $p$ and $q$, well fitted by a linear dependence (line) on the shape factor $\xi = (q-p)/p$. Lower inset: unit cell geometry labeled with graphene-like sublattices A and B. **b**, Landau gauge potential $A_x(y) = -B_z y$ synthesized by a SC with a linearly varying $\xi$ along the $y$ direction but a translational invariance along the $x$ direction. **c**, Spectrum near the Dirac frequency, simulated for a 91-layer gradient SC in a ribbon geometry with $B_z = -0.013 a^{-2}$. Bold lines: analytically predicted LLs. **d**, Pressure amplitude distributions for the three eigenstates labeled in **c**, exhibiting features of bulk ($s_2$) and edge ($s_1$ and $s_3$) states. Insets: Local views of the strong field regions, revealing a sublattice polarization of the bulk state, in contrast to those of edge states.

We consider first a 2D SC arranged by a triangular lattice of uniform circular scatterers in air background. The scatterers are acoustically rigid and the air channels surrounding them form a graphene-like honeycomb lattice, in which the two sublattices are labeled by A and B. Such acoustic graphene hosts Dirac cones at $K$ and $K'$ points, around which the effective Dirac model applies. As the scatterers are deformed into ellipses, the crystal symmetry is reduced from $C_{3v}$ to $C_{2v}$ and the Dirac points move along the mirror line of the hexagonal Brillouin zone (Fig. 1a). This gives a constant vector potential $\mathbf{A} = \delta\mathbf{k} = (\delta k_x, 0)$ for a uniformly deformed SC. Here $\hbar/e = 1$ is used for our classical system[13-17]. Interestingly, the momentum shift can be captured by a simple relation $\delta k_x = c_\gamma \xi$ (Methods), where the dimensionless factor $\xi = (q-p)/p$ characterizes the shape deviation from a perfect circle, and the coefficient $c_\gamma$ depends on the spatial filling ratio of the scatterer, $\gamma$. Figure 1a exemplifies the effectiveness of this intuitive linear relation at $\gamma = 0.47$ for a large range of $\delta k_x$, associated with $c_\gamma = 0.59\pi/a$ and the lattice constant $a = 2.5$ cm used hereafter. This enables an easy design to attain Landau gauge $A_x(y) = -B_z y$ by linearly varying the shape factors of scatterers along the $y$ direction such: $\xi(y) = -B_z y / c_\gamma$ (Fig. 1b). Here the constant $B_z$ defines the strength of acoustic PMF orientated in the $z$ direction. Physically, the PMF comes from the spatial modulation of the effective couplings among the lattice sites through tailoring the air-channel geometry of the SC. As a result, the linear Dirac cones quantize into a sequence of discrete Landau Levels (LLs) $\Delta_n = \text{sign}(n)\sqrt{|n|}\omega_c$ measured from the Dirac frequency $\omega_D$, where $\omega_c = v_D\sqrt{2|B_z|}$ is an acoustic analog of the cyclotron frequency[3,7,10,11]. Such a LL structure is unique to the massless Dirac bands, in sharp contrast to the equally spaced LLs of the conventional quadratic bands[10,11].

Our design strategy can be exemplified by a 91-layer gradient SC ribbon with PMF $B_z = -0.013a^{-2}$. Clearly, its spectrum (Fig. 1c) exhibits discrete flat levels centered by the Dirac frequency $\omega_D = 6.97$ kHz, as a hallmark of Landau quantization for massless Dirac systems. The frequencies of the first few LLs match perfectly with the analytical prediction. The states in each LL are accidentally degenerate, as they share the same energy but their wave functions center at different $y$-coordinates[10,11]. Near a ribbon edge[26], the acoustic cyclotron orbits become open, and their frequencies deviate gradually from the LL values as these states become more and more localized at the edge. Here we focus more on the zeroth LL, which is pinned to the Dirac frequency regardless of the (pseudo)magnetic field strength. Figure 1d shows the mode profiles for one bulk LL state and two edge states. Evidently, the bulk-state vibrations behave strikingly different at the two sublattices: the field concentrates around the A sublattice but vanishes at the B sublattice. This is a direct consequence of the graphene-like Dirac systems, where the zeroth LL states are sublattice polarized[10-12], in contrast to the states in other LLs that occupy both sublattices. The polarization weakens as the state departs from the Landau plateau due to the interaction with the physical boundary (see insets).

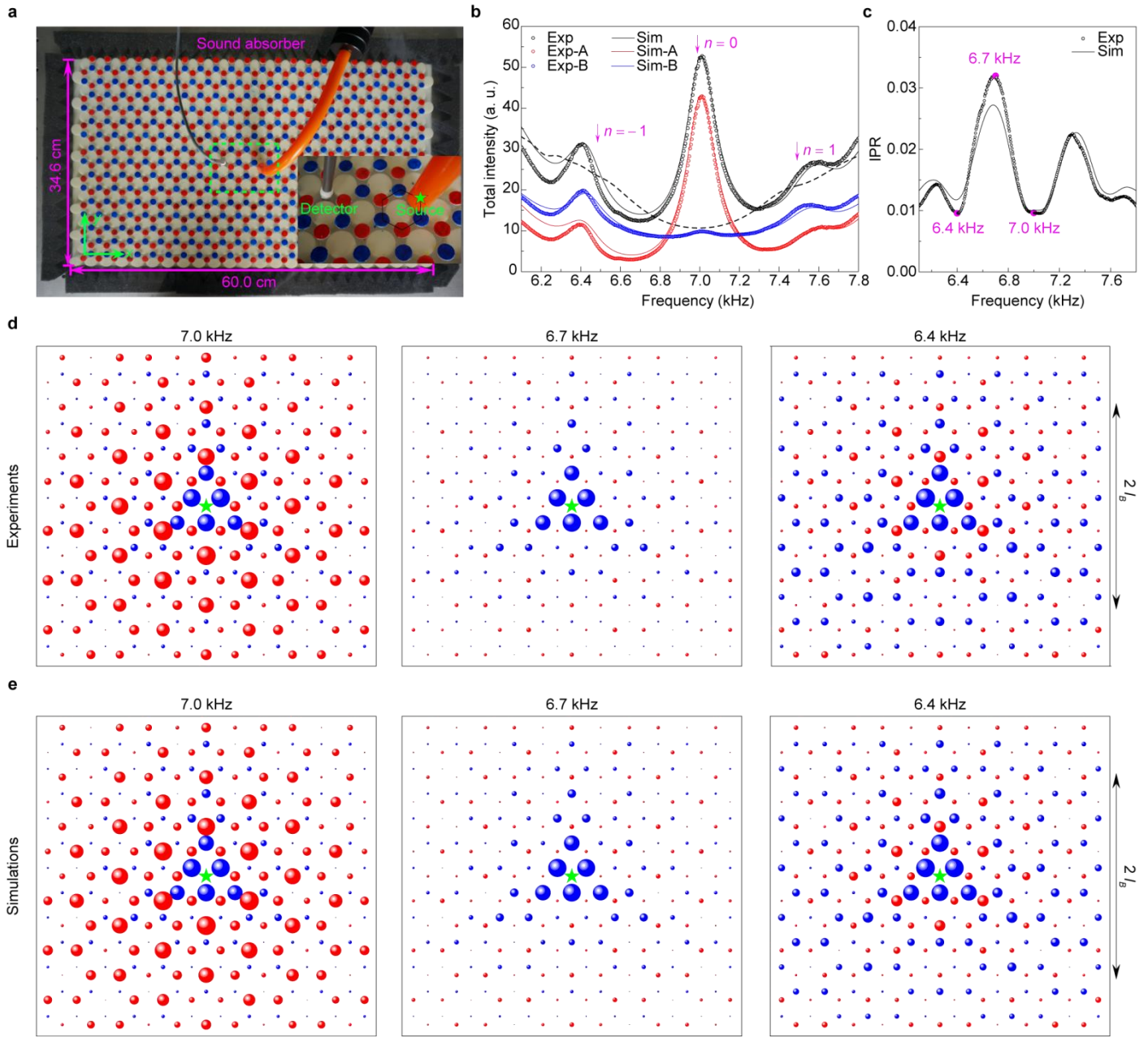

**Figure 2 | Experimental detection of the acoustic Landau quantization. a**, Experimental setup, with a strong PMF ($B_z = -0.078a^{-2}$) produced in a 16-layer SC. Inset: a zoom-in image which indicates the sound source and detector. **b**, Frequency spectra for the total sound energy plotted around the lowest three LLs, excited by the sound source located at one A-site near the sample center (the star in **d**). The experimental result (black open circles) agrees well with the numerical one (black solid line), whose peak structures follow the model prediction (purple arrows). The spectrum simulated for a uniform SC with $\xi = 0$ (black dashed line) is provided for comparison. The individual sound energies for the A and B sublattices confirm the polarization nature of the zeroth LL. **c**, Measured and simulated inverse participation ratio (IPR) curves.

A larger IPR value reveals a more localized state. **d**, Direct visualization of the measured pressure distributions for the three frequencies labeled in **c**. The field amplitudes at the A and B sublattices are characterized by the sizes of the red and blue spheres, respectively. $l_B$ is the effective magnetic length indicating the size of cyclotron orbit of each acoustic LL state. **e**, Numerical comparisons for **d**.

The acoustic Landau quantization can be detected experimentally through the peak structure of the excitation spectrum, which reflects the high density of states of the discrete LLs. Different from Fig. 1c, here we consider a 16-layer SC with a much stronger acoustic PMF, $B_z = -0.078 a^{-2}$, which is equivalent to ~840 Tesla for graphene under a real magnetic field. This enables a much larger cyclotron frequency $\omega_c = v_D \sqrt{2|B_z|}$ and thus much wider LL gaps, which greatly facilitate our observation of the acoustic LLs and the associated quantum-Hall-like edge modes. Figure 2a shows our experimental setup. The sample was precisely fabricated with photosensitive resin via 3D printing technique; it is integrated with 376 elliptical scatterers on a substrate and covered by an optically transparent holey plate to form a 2D sound waveguide. To reveal the sublattice polarization nature of the zeroth LL state, a point-like sound source is located at one A-site near the sample center. The plugs that seal the holes, colored in red and blue respectively for A and B sublattices, will be opened one-by-one for sound detection.

Figure 2b presents the measured frequency spectra (black circles) for the total sound energy summed over the A and B sublattices. It shows clear quantization peaks created by the acoustic PMF, in sharp contrast to the quadratic background spectrum (black dashed line) simulated for a 16-layer sample with zero PMF. Specifically, the primary peak centered at the Dirac frequency (~7.0 kHz) is responsible for the zeroth LL, which is a key feature of Landau quantization in a relativistic Dirac system. The secondary peaks centered at ~6.4 kHz and ~7.5 kHz correspond to the $n=-1$ and $n=1$ LLs, respectively. All peak positions are captured well by the effective Dirac model under Landau gauge (purple arrows). To confirm the polarization nature of the zeroth LL state, Fig. 2b also shows the individual excitation spectra for A and B sublattices. Clearly, the A-sublattice spectrum (red circles) exhibits a pronounced peak around 7.0 kHz. This is in sharp contrast to the B-sublattice spectrum (blue circles), in which the remnant value comes from several B-sites immediately surrounding the sound source (see Fig. 2d, left panel). Physically, the sublattice polarization stems from the broken inversion symmetry of the gradient SC. As expected, such strong polarization feature is not exhibited by the $n=\pm 1$ LLs, for which the peaks (despite weak) emerge simultaneously for the A and B sublattices. All experimental data (circles) are reproduced quantitatively by our full-wave simulations (solid lines) with appropriate absorption taken into account.

To disclose the localization features of LL states, an inverse participation ratio (IPR) function[17] $\left(\sum_{i=A,B}|p_i|^4\right)\left(\sum_{i=A,B}|p_i|^2\right)^{-2}$ that sums over all the A and B sites is calculated based on our data. Essentially, this function characterizes the localization degree of a state: the larger the IPR value, the stronger the localization. Consistently, both the measured and simulated IPR curves in Fig. 2c exhibit three dips responsible for the lowest three LLs and two pronounced peaks for the LL gaps in between. Benefited from the macroscopic controllability of our airborne sound experiments, the localization features can also be directly visualized through the excited field patterns. As examples, Fig. 2d shows the experiment data for three typical frequencies. At the zeroth LL frequency (7.0 kHz), the B-sublattice amplitude (blue spheres) attenuates quickly except for the few sites surrounding the sound source, whereas the A-sublattice amplitude (red spheres) spreads over a length scale of the effective magnetic length $l_B = |B_z|^{-1/2} = 3.6a$, as a manifestation of the peculiar sublattice polarization again. Note that both $K$ and $K'$ valleys are excited simultaneously, and that their interference leads to an intriguing field pattern of the A sublattice. (The sublattice polarization is the same for the two valleys because of time-reversal symmetry.) At the $n = -1$ LL frequency (6.4 kHz), both the A and B sublattices are excited as expected. At the frequency in between (6.7 kHz), however, the amplitudes of both sublattices are suppressed, as evidence for the gap states due to the Landau quantization. Again, all the experimental data are highly consistent with those simulations in Fig. 2e. For better contrast, similar data for the B-site excitation are provided in Extended Data; in that case, the zeroth LL states are hardly excited due to the A-sublattice polarization.

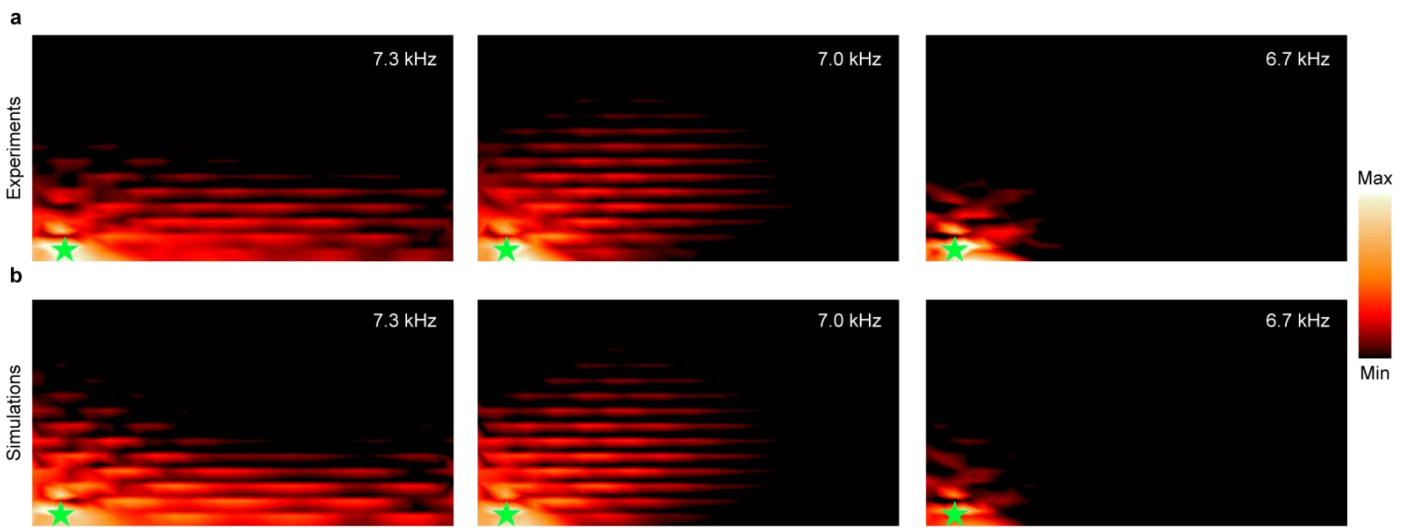

**Figure 3 | Acoustic quantum-Hall-like edge states generated by the effective magnetic field. a,b**, The measured and simulated pressure amplitude distributions at three different frequencies. In each case the

green star indicates the position of the sound source. Note that the excitations are no longer sensitive to the detailed source position since the edge states are less polarized (Fig. 1d). Our experiments reproduce well the simulations.

Dispersive and propagating quantum-Hall-like edge modes can be observed in the gaps between two neighboring acoustic LLs. To the best of our knowledge, to date such PMF-induced edge states have not been detected in graphene, mainly due to the technical challenge in creating the uniform PMF over a large area[5,6]. Here we focus on the edge states that propagate along the bottom boundary, in which the sound leakage is prevented by attaching an acoustically rigid strip. As shown in Fig. 3a, the sound source is positioned at the bottom left of the sample to selectively excite the edge modes that travel rightwards. (If the sound source is positioned at the bottom-right of the sample, leftward propagating edge states will be excited since the PMF does not break time-reversal symmetry.) For comparison, we display the measured pressure distributions for three typical frequencies. Similar to the LL structure in Fig. 1c, the frequency 7.3 kHz falls into the gap between the $n=0$ and $n=1$ LLs, and a well-confined edge state emerges in the bottom boundary. By contrast, at 7.0 kHz (the frequency for $n=0$ LL) more sound energy spreads into the bulk, and at 6.7 kHz (between the $n=-1$ and $n=0$ LLs) the pressure profile becomes much more localized around the source. If the sound source is positioned at the upper boundary, on the contrary, a propagating edge state appears at 6.7 kHz whereas the field pattern becomes localized at 7.3 kHz (Extended Data). All the experimental data (Fig. 3a) agree well with our numerical calculations (Fig. 3b).

In conclusion, we have achieved the celebrated relativistic Landau quantization in gradient acoustic graphene and revealed its previously elusive quantum-Hall-like edge states generated by the PMF. Our design for attaining the uniform PMF is simple: only one geometric parameter is tailored along one direction. As such, we can achieve a giant PMF without misshaping the honeycomb lattice. This is markedly different from those realized in electronic[5,6,12] and photonic[13] systems, in which the PMF is introduced by applying a triaxial strain to deform the lattice. In addition, our experiments are performed in a real 2D system, which allows a direct demonstration of the frequency-resolved LL spectra and associated quantum-Hall-like edge states. This is much different from the strained photonic graphene realized by dielectric cylinder arrays[13], in which the measurements are made at a single frequency and the energy is mimicked by the out-of-plane propagation constant that is difficult to resolve. Our scheme can be easily extended to other artificial structures such as a patterned photonic crystal slab on a silicon chip, and the high density of states of the acoustic LLs provides a novel mechanism for enhancing sound emission and nonlinear wave mixing.

**Acknowledgements** This work is supported by the National Basic Research Program of China (Grant No. 2015CB755500); National Natural Science Foundation of China (Grant Nos. 11774275, 11674250, 11534013, 11747310); Natural Science Foundation of Hubei Province (Grant No. 2017CFA042). FZ was supported by the UT-Dallas research enhancement funds.